\documentclass[aps,prl,longbibliography,twocolumn,groundaddress,floatfix]{revtex4-2}
\usepackage{amsmath,amssymb,bm,graphicx,color,gensymb,bbold,hyperref,keyval,url,latexsym}
\usepackage{xcolor}
\usepackage[normalem]{ulem} 
\usepackage{CJK}
\usepackage{float}

\begin{document}
\title{Quantum Phases in the Honeycomb-Lattice $J_1$--$J_3$ Ferro-Antiferromagnetic Model}

\begin{CJK*}{UTF8}{}
\author{Shengtao Jiang (\CJKfamily{gbsn}蒋晟韬)}
\author{Steven R. White}
\author{A. L. Chernyshev}
\affiliation{Department of Physics and Astronomy, University of California, Irvine, California 92697, USA}
\date{\today}

\begin{abstract}
Using  large-scale density-matrix renormalization group calculations and minimally augmented spin-wave theory, we demonstrate that the phase diagram of the  quantum  $S\!=\!\frac12$ $J_1$--$J_3$ ferro-antiferromagnetic  model on the honeycomb lattice differs dramatically from the classical one. It hosts the double-zigzag and Ising-z phases as unexpected intermediaries between ferromagnetic and zigzag states that are also extended  beyond their classical regions of stability. In broad agreement with quantum order-by-disorder arguments, these collinear phases replace the classical spiral state.
\end{abstract}
\maketitle
\end{CJK*}

{\it Introduction.}---Ever since the Anderson's seminal work  on the resonating valence-bond state~\cite{ANDERSON1973}, the significant role that can be played by quantum fluctuations in magnets with competing interactions has remained at the forefront of condensed matter  physics, inspiring a multitude of quests for exotic states, models that can realize them, and real materials that can host them~\cite{balents_2010,Norman_16,Kanoda_17,Knolle_19,KITAEV,Becca_01}. The elusive spin-liquid states with strongly entangled spins are but one example~\cite{balents_2010}; others include valence-bond phases with spatial symmetry breaking~\cite{Batista_14,Mila_06,plaquette-j1j2square,Mila_02,Mila_13a,Gaulin_SS_12,Becca_17}, quantum multipolar spin nematics  that are quantum analogues of  liquid crystals~\cite{nematic-our,nematic-mike,PencLauchli2011,nematic-exp}, and an especially extensive class of  unconventional magnetically {\it ordered} phases that do not appear in the classical solutions of the underlying spin models~\cite{Henley_87,Henley_89,starykh2015,Gingras_13,RossPRX_11,Rau_review_pyro19,Hallas_review18,%
Rau_19,Rau_obd,MZh_fcc_22}. 
It is the latter group of phenomena that creates a broader context for the present study.

The ordered phases that are not favored classically but are stabilized in the quantum $S\!=\!\frac12$ limit have attracted  significant attention in the search for Kitaev magnets on the honeycomb lattice~\cite{Witczak_Krempa14,YBK_2016,rau_spin-orbit_2016,Winter_17}. Recently, this extensive experimental and theoretical effort has expanded to  the Co$^{2+}$ materials~%
\cite{Khaliullin18,Khaliullin20,Armitage23,Cava20,Ross18,Lorenz23,Park22,Lynn23,Songvilay20,Zaliznyak22,
Regnault_18,arun-firstprinciple,pavel-dZZ,Broholm_2023}.
It appears that the minimal $XXZ$-anisotropic $J_1$--$J_3$ model with  ``mixed'' ferro-antiferromagnetic (FM-AFM) couplings, given by
\vskip -0.15cm
\noindent
\begin{equation}
    \label{eq:j1j3mod}
    H=\sum_{n=1,3}\sum_{\langle ij \rangle_n} J_n\Big(S^x_iS^x_j+S^y_iS^y_j+\Delta_n S^z_iS^z_j\Big),
\end{equation}
\vskip -0.25cm
\noindent
provides a tantalizingly close description for many of these compounds~\cite{Rethinking,dejongh1990,usLPR,Regnault_18,arun-firstprinciple,pavel-dZZ,Broholm_2023}, calling for its unbiased study.
Here $\langle ij \rangle_{1(3)}$ stands for the first-(third-)neighbor bonds, $J_1\!=\!-1$ is  the energy unit,   $J_3\!>\!0$, and  $0\!\leq\!\Delta_n\!\leq\!1$ are the $XXZ$ anisotropies. 
We note that earlier pre-Kitaev searches for exotic quantum states have focused on a pure AFM $J_1$--$J_2$--$J_3$ honeycomb-lattice model~\cite{Oitmaa91,Singh99,j1j2j3-ed2001,j1j2-arun,Oitmaa11,Cabra11,j1j2-rigol2,j1j2-bishop,j1j2-gong,j1j2-steve1,j1j2-steve2,j1j2-ganesh}, motivated by the expectation of  stronger fluctuations due to the lattice's low  coordination number  and  by the degeneracies in its  classical phase diagram~\cite{j1j2-arun}. 

\begin{figure}[t]
\centering    
\includegraphics[width=1.0\columnwidth]{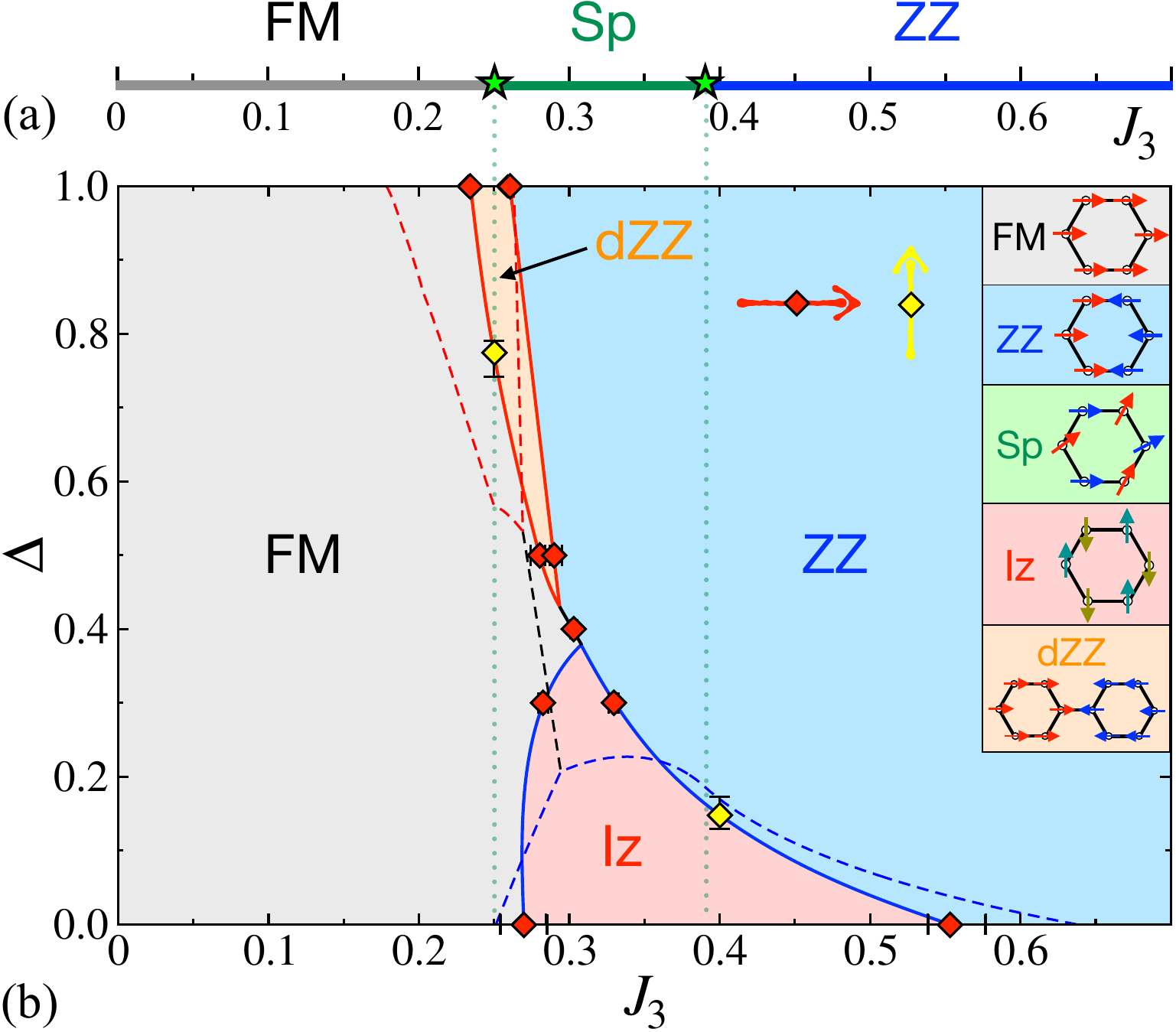}
\vskip -0.35cm
\caption{The classical (a)  and  quantum (b) phase diagrams of the $XXZ$ $J^\Delta_1$--$J_3$ model (\ref{eq:j1j3mod}) with the ferromagnetic (FM), zigzag (ZZ), spiral (Sp), double-zigzag (dZZ), and  Ising-$z$ (Iz) phases. The solid lines are phase boundaries interpolating  transition points (diamonds) inferred from the DMRG scans  along $J_3$ (red) and $\Delta$ (yellow).  The vertical and dashed lines are classical and MAGSWT phase boundaries, respectively. Spins are in-plane for all  phases except Iz, see also Fig.~\ref{fig:scanhori}.}
\label{fig:phd}
\vskip -0.6cm
\end{figure}

The model (\ref{eq:j1j3mod}) was studied in the 1970s~\cite{Rastelli}, yielding the classical phase diagram reproduced in Fig.~\ref{fig:phd}(a). These phases are independent of $\Delta_n$ because  all relevant classical states  are coplanar. The ground state is FM for small $J_3$,  while zigzag (ZZ) order is preferred for large $J_3$, and the ferrimagnetic  spiral phase (Sp)  continuously interpolates between  FM and ZZ.

In this Letter, we combine  density-matrix renormalization group (DMRG) and minimally-augmented  spin-wave theory (MAGSWT)  to obtain the groundstate phase diagram of the quantum $S\!=\!\frac12$ model (\ref{eq:j1j3mod}). We focus on the {\it partial} $XXZ$ version of the model (\ref{eq:j1j3mod}), with the $J_3$-term left in the Heisenberg limit, $\Delta_3\!=\!1$, referred to as the $J_1^\Delta$--$J_3$ model. This choice is motivated by real materials, in which further exchanges tend to be more isotropic~\cite{valenti16,Winter_17}.
The standard version of the model with equal anisotropies, $\Delta_1\!=\!\Delta_3$, referred to as the {\it full} $XXZ$ or  $J_1^\Delta$--$J_3^\Delta$ model, is considered too.

\begin{figure*}[t]
\centering    
\includegraphics[width=2.0\columnwidth]{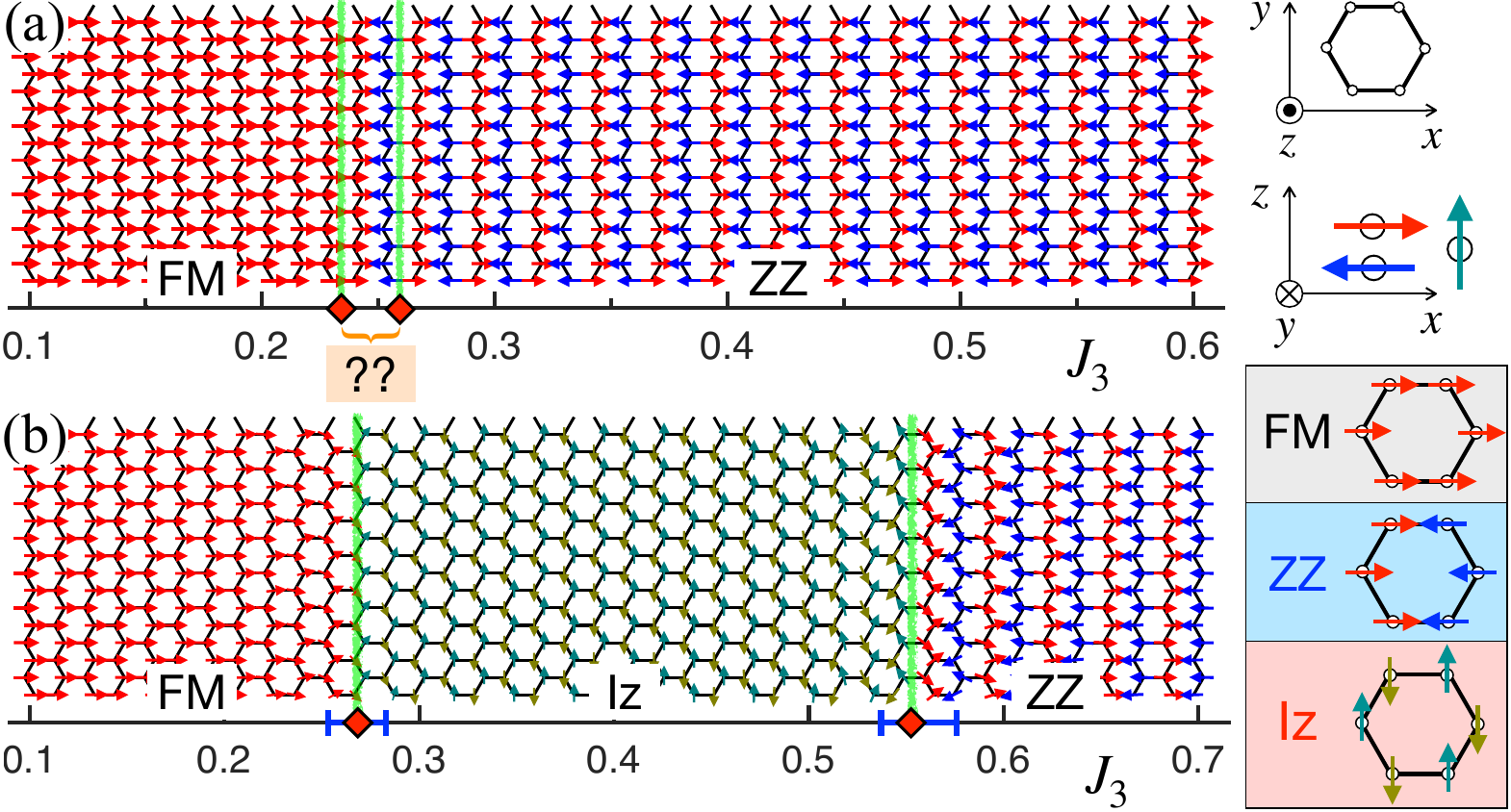}
\vskip -0.2cm
\caption{Long-cylinder scans of the $J_1^\Delta$--$J_3$ model (\ref{eq:j1j3mod}) vs $J_3$ in the  (a) Heisenberg ($\Delta\!=\!1$) and (b) $XY$ ($\Delta\!=\!0$) limit. The arrows show the local ordered moment $\langle {\bf S}_i\rangle$. FM, ZZ, and Iz phases are indicated and transitions are determined as described in text. The honeycomb lattice is in the $xy$ plane while spins shown in the figure are in the $xz$ plane.}
\label{fig:scanhori}
\vskip -0.5cm
\end{figure*}

{\it Phase diagram.}---%
Our phase diagram for the  $S\!=\!\frac12$  $J_1^\Delta$--$J_3$ model is given in Fig.~\ref{fig:phd}(b). In a dramatic deviation from the classical case, we find  two unconventional phases stabilized by quantum fluctuations---the double-zigzag (dZZ)  and Ising-$z$ (Iz) phases---as  intermediary between the FM and ZZ phases. The  FM and ZZ phases also extend well beyond their classical regions  to completely supersede  the non-collinear classical spiral phase.

 The solid lines are  phase boundaries  interpolating transition points obtained from the DMRG  long-cylinder DMRG ``scans'' by varying $J_3$  or $\Delta$, as well as from the more precise measurements. The dashed lines are phase boundaries of the same phases obtained by MAGSWT, with  both approaches described below.  

The qualitative agreement between these approaches is quite remarkable. Both methods produce the classically unstable dZZ and Iz phases, both expand the FM and ZZ phases beyond their classical ranges, and both eliminate the Sp phase. These findings are also in a broad agreement with  order-by-disorder arguments~\cite{Henley_89,Rau_obd}, which generally favor  collinear phases.

We note that  recent studies of  related models also found the Sp phase to be absent~\cite{Arun2022,Trebst2022}.  However, our conclusions on the nature and extent of the quantum phases that replace it differ substantially from theirs. For the details on these differences for the $J_1$--$J_3$ and other models, see Ref.~\cite{sm} and the discussions below.

The $U(1)$-preserving Iz phase, with  spins ordered N\'{e}el-like along the $z$ axis, has been first discovered in the $XY$ $J_1$-$J_2$  AFM-AFM model~\cite{j1j2-steve2}, where Iz order is stabilized solely by quantum effects with no exchange coupling favoring it. In our case, we find the $z$ axis component of the $J_3$-exchange in the $J_1^\Delta$-$J_3$ model  crucial for stabilizing the Iz phase in a wide range of parameters, see Fig.~\ref{fig:phd}(b). In contrast to Ref.~\cite{Trebst2022}, we find only a very narrow Iz phase in the $J_1^\Delta$-$J_3^\Delta$ model. The spin-liquid phases in this model \cite{Arun2022,Trebst2022} are also not supported \cite{sm}.

 The dZZ phase has been recently reported experimentally~\cite{Regnault_18} and  found  favored  by the {\it bond-dependent} extensions of the $XY$ $J_1^\Delta$--$J_3^\Delta$ model~\cite{pavel-dZZ,Broholm_2023}. Instead, we find the dZZ phase  already in the Heisenberg limit of the  principal $J_1$--$J_3$ model (\ref{eq:j1j3mod}), see Fig.~\ref{fig:phd}(b). 

{\it DMRG calculations.}---%
DMRG calculations were performed on the $L_x\!\times\!L_y$-site honeycomb-lattice open cylinders of width $L_y$ up to 16 (8 honeycomb cells), using the ITensor library~\cite{itensor}. The majority of the results were obtained on the so-called X-cylinders (XC)~\cite{j1j2-steve1}, in which the first-neighbor bond is horizontal, while both X- and Y-cylinders (YC) were used for more  delicate phases~\footnote{We typically perform 16 sweeps and reach a maximum bond dimension of $m\!\sim\!3000$ to ensure good convergence with the truncation error of $\mathcal{O}(10^{-5})$.}. We allow for a spontaneous breaking of the spin $U(1)$ symmetry~\footnote{Such symmetry breaking in DMRG mimics the 2D system, see Sec.~I of the SI in Ref.~\cite{tt'j}.}, enabling us to measure the local ordered moment $\langle {\bf S}_i\rangle$ instead of the correlation function.

Our main exploratory tool is the long-cylinder ``scans,'' in which  one parameter, $J_3$ or $\Delta$, is varied along the length of the cylinder with $L_x$ up to 40. It provides 1D cuts through the 2D phase diagram~\cite{scan1,scan3,scan4,scan5}, see Fig.~\ref{fig:scanhori}, which give approximate phase boundaries. By narrowing parameter ranges of the scans  one can determine the boundaries with  increased precision,  distinguish  first- and second-order transitions~\cite{nematic-our},  and  uncover hidden phases. In  cases when  the  phase boundary is less obvious, we  utilize the fixed parameter (non-scan) calculations  on  clusters up to $16\!\times\!16$, with the aspect ratio that closely approximates the 2D  thermodynamic limit~\cite{FS}. 

In Fig.~\ref{fig:scanhori}, we present two long-cylinder scans for the $J_1^\Delta$--$J_3$  model (\ref{eq:j1j3mod}), one in the Heisenberg limit, $\Delta\!=\!1$, and the other in the $XY$ limit, $\Delta\!=\!0$, vs $J_3$. In the Heisenberg limit, Fig.~\ref{fig:scanhori}(a), the transition from FM to ZZ is very sharp and FM phase seems to terminate right at the classical boundary of this state, $J^{cl}_3\!=\!0.25$. However, one would expect that the FM phase should retreat from this boundary, as the competing ZZ state is fluctuating in the Heisenberg limit, while the FM state is exact.  The subsequent analysis reveals a hidden intermediate dZZ state, discussed next. We note that the scan calculation in Fig.~\ref{fig:scanhori}(a) misses it not only due to the narrow region of the dZZ phase, but also because of the high symmetry of the model in the Heisenberg limit, which requires additional effort to avoid metastable states. 

Fig.~\ref{fig:scanhori}(b) for the $XY$ limit shows transitions from the FM to Iz and from Iz to ZZ vs $J_3$. By using scans in the narrower ranges of $J_3$, we verify that the spiral-like spin patterns in the transition regions in Fig.~\ref{fig:scanhori}(b)  are proximity effects of the neighboring phases, not additional phases.  The phase boundaries shown in Fig.~\ref{fig:scanhori}(b) and used in the phase diagram in Fig.~\ref{fig:phd}(b) are the crossing points of the order parameters   vs $J_3$~\cite{sm}. 
The error bars are the width of the transition region in the scans, where a discontinuous transition is assigned a width equal to the parameter change over  one lattice spacing.

\begin{figure}[t]
\centering    
\includegraphics[width=1.0\columnwidth]{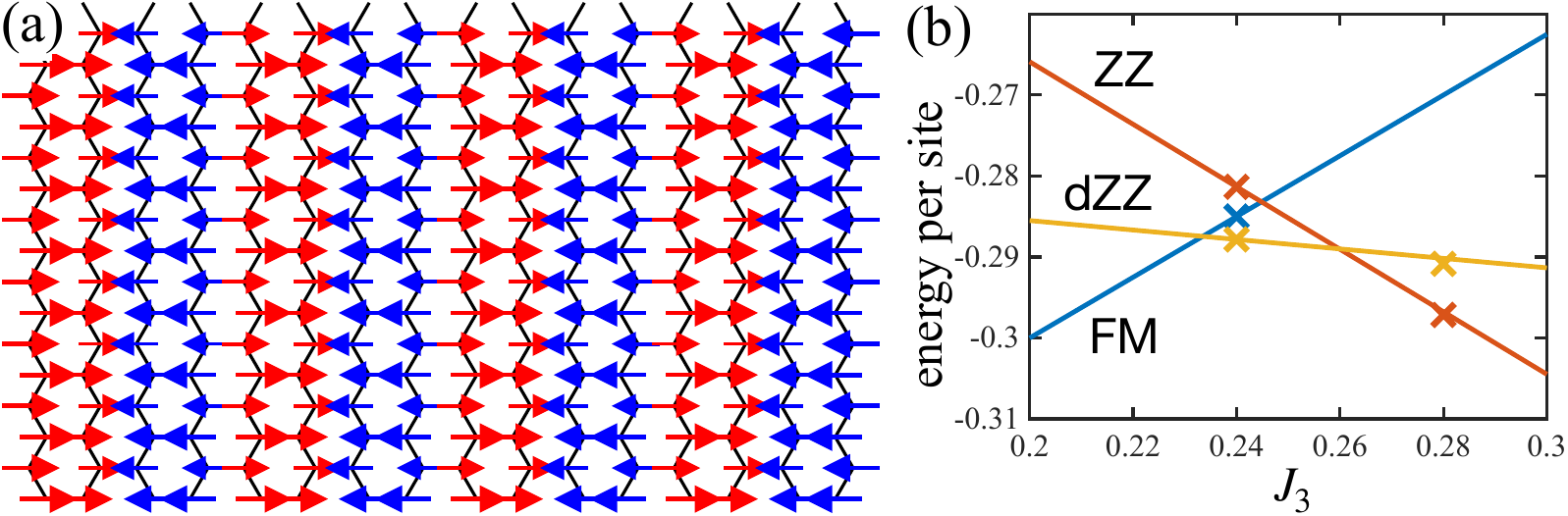}
\vskip -0.3cm
\caption{(a) Ordered moments in the $16\!\times\!16$ non-scan  cluster for $J_3$=0.24, showing dZZ pattern. (b) Energies of the three competing phases vs $J_3$, crosses are DMRG results and higher-energy states are metastable. Lines are extrapolated energies, $\langle \psi_i|H(J_3)|\psi_i\rangle$, where $\psi_i$ are the three states at $J_3\!=\!0.24$.}
\label{fig:dzz}
\vskip -0.5cm
\end{figure}

In the Heisenberg limit, the three states, FM, dZZ, and ZZ, compete in the proximity of  the classical FM boundary $J_3\!=\!0.25$.  Because of the high spin-symmetry of the model, and depending on the initial state, all three  can be stabilized in the non-scan DMRG simulations, such as the one shown in Fig.~\ref{fig:dzz}(a)  for $J_3\!=\!0.24$  in the $16\!\times\!16$ cluster. As is shown in Fig.~\ref{fig:dzz}(b), the energy of the dZZ is  the lowest, with the FM and ZZ being metastable, suggesting that the transitions between the corresponding phases are   first order. To identify their phase boundaries, we  compare the energies of these three states as a function of $J_3$ using extrapolations based on the spin-spin correlations  extracted at $J_3\!=\!0.24$ from the center of the cluster for each of the states.  While the FM line is exact in this limit, the extrapolated energies for ZZ and dZZ  are also very close to the ones given by a direct DMRG calculation at a different value of $J_3$, justifying the analysis, see Fig.~\ref{fig:dzz}(b). The  dZZ phase is found to be confined between $J_3$=0.2333 and 0.2596.

\begin{figure}[t]
\centering    
\includegraphics[width=1.0\columnwidth]{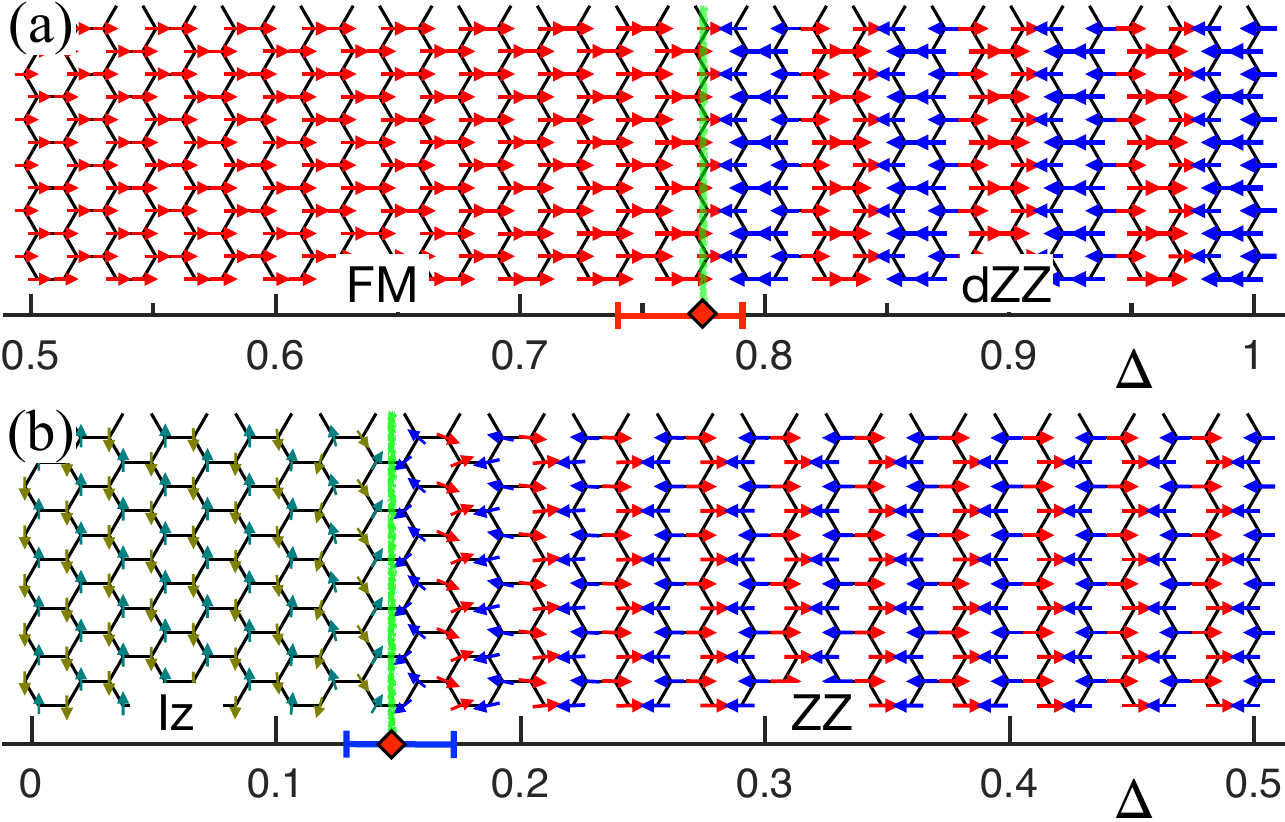}
\vskip -0.3cm
\caption{Long-cylinder $\Delta$-scans of the $J_1^\Delta$--$J_3$ model (\ref{eq:j1j3mod}) for  (a) $J_3\!=\!0.25$ and (b) $J_3\!=\!0.4$. Notations are as in Fig.~\ref{fig:scanhori}.}
\label{fig:scanvert}
\vskip -0.6cm
\end{figure}

The lower spin-symmetry away from the Heisenberg limit  helps to reveal the dZZ phase more readily, see Fig.~\ref{fig:scanvert}(a) for a long-cylinder scan along the $\Delta$ axis  and fixed $J_3\!=\!0.25$, confirming  the presence  of this phase in an extended region of the phase diagram in Fig.~\ref{fig:phd}. A similar $\Delta$-scan for $J_3\!=\!0.4$ in Fig.~\ref{fig:scanvert}(b) compliments the $J_3$-scans in establishing  boundaries of the Iz phase. 

By using a combination of the narrower  ranges of the scans and fixed-parameter non-scans, we find that the dZZ phase persists somewhat below $\Delta\!=\!0.5$ while the Iz phase ends close to  $\Delta\!=\!0.4$, where the FM-to-ZZ transition appears to be direct,  see Fig.~\ref{fig:phd} and \cite{sm}. Although we cannot completely rule out the Iz state  for $\Delta\!=\!0.4$,  it must be extremely narrow if it exists.  

{\it Minimally-augmented spin-wave theory.}---%
The standard SWT is   successful at accounting for  quantum effects in the ordered states \cite{RMP_13}, but cannot describe either the ordered phases that are not classically stable, or the  shifts of the phase boundaries by  quantum fluctuations. An analytical approach to address this problem, originally proposed for the classically unstable field-induced states in the transverse-field Ising and frustrated Heisenberg models \cite{Mila_12,Mila_14,Mila_13}, can be successfully applied here.

The method consists of introducing a  local field in the direction of the ordered moment ${\bf n}_i$ for the proposed (unstable) classical spin configurations, leading to a shift of the chemical potential in the bosonic SWT language
\vskip -0.15cm
\noindent
\begin{equation}
\label{eq_dH}
\delta{\cal H}=\mu\sum_{i} \left(S-{\bf S}_i\cdot{\bf n}_i\right)=\mu\sum_{i} a^\dag_i a_i,
\end{equation}
\vskip -0.25cm
\noindent
while leaving the classical energy of the state unchanged. The {\it minimal} value of  $\mu$ is chosen to ensure stability of the spectrum, i.e., that the squares of all eigenvalues of the SWT matrix are positive definite. Then, the  energy of the proposed spin state, ${\cal E}\!=\!E_{cl}+\delta E$, with the $1/S$-correction to the groundstate energy $\delta E$, is well-defined and can be compared with the energies of the competing states calculated to the same  $O(S)$ order. 

The power of the method, coined as the {\it minimally augmented SWT} (MAGSWT),  is not only in its simplicity, but in the form of Eq.~(\ref{eq_dH}), which guarantees that its contribution to the Hamiltonian is positive for $\mu\!>\!0$. In turn, this implies that the so-obtained groundstate energy ${\cal E}$ is an {\it upper bound} for the energy of the suggested spin state to the order $O(S)$. This method allows one to consider the phase beyond its classical range of stability and inspect states that are classically not competitive, but can lower their energy due to quantum fluctuations. The new phase boundaries are determined from the crossings of the energies ${\cal E}$ for the competing phases as a function of the varied parameter(s). 

We note that MAGSWT may not be applied to an arbitrary classically-unstable state~\cite{Mila_13}, 
with the absence of the linear-bosonic terms in the $1/S$-expansion for a given state being a sufficient criterion of its applicability. 

\begin{figure}[t]
\centering    
\includegraphics[width=1.0\columnwidth]{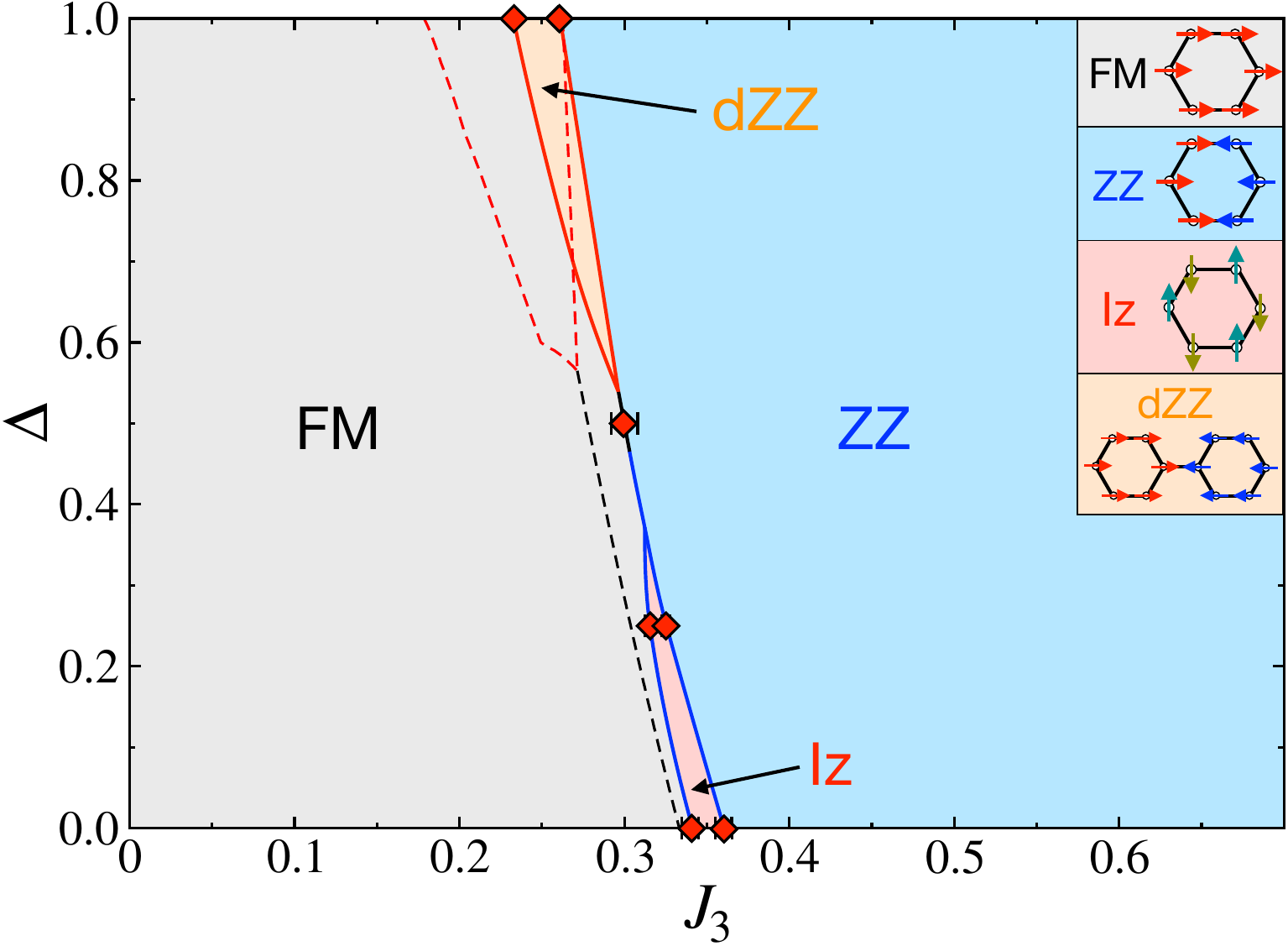}
\vskip -0.35cm
\caption{The  quantum $S\!=\!\frac12$ phase diagrams of the full $XXZ$ $J^\Delta_1$--$J^\Delta_3$ model (\ref{eq:j1j3mod}), c.f. Fig.~\ref{fig:phd}(b). See text.}
\label{fig:phd_full}
\vskip -0.6cm
\end{figure}

{\it MAGSWT results.}---%
In  case of the $XXZ$ $J_1^\Delta$--$J_3$ model (\ref{eq:j1j3mod}), all four competing phases of interest are collinear, which guarantees the absence of the linear-bosonic  terms, while the non-collinear Sp state  is not the subject of MAGSWT, as it corresponds to a minimum of the classical energy in its entire possible range of existence. 

The technical procedure of extracting minimal $\mu$ vs $J_3$ and $\Delta$ for each phase is discussed in Ref.~\cite{sm}. We note that the limiting $XY$ and Heisenberg cases and select momenta are useful for obtaining  analytical expressions for  $\mu(J_3,\Delta)$, eliminating the need of a numerical scan of the   momentum space for spectrum instabilities. With that, the energy surfaces ${\cal E}(J_3,\Delta)$ are readily obtained for each phase and the MAGSWT phase boundaries are drawn from the intersections of such surfaces. 

The resulting phase boundaries are shown in Fig.~\ref{fig:phd}(b) by the dashed lines. Most, if not all, of the features already discussed above are present. The noncollinear Sp phase  is not effective at benefiting from  quantum fluctuations, in agreement with the  order-by-disorder arguments~\cite{Henley_89}, and is wiped out. The classically-unstable dZZ and Iz phases are extensive and both FM and ZZ expand beyond their classical borders. A  close quantitative agreement with the DMRG phase boundaries can also be observed, with most discrepancies concerning the borders of the less-fluctuating FM phase \cite{sm}. 
Otherwise, the entire picture for the $J_1^\Delta$--$J_3$ model in Fig.~\ref{fig:phd}(b) is in  rather astonishing agreement with the numerical data.

{\it The $J_1^\Delta$--$J_3^\Delta$ model.}---%
The phase diagram of the full $XXZ$ model (\ref{eq:j1j3mod}) with equal anisotropies in both terms, obtained using the same methods as described above, is presented in Fig.~\ref{fig:phd_full}. It repeats most of the trends of the partial $XXZ$ model  in Fig.~\ref{fig:phd}(b), such as the absence of the Sp phase, expansion of the FM and ZZ, and the presence of the two unconventional phases, Iz and dZZ.  

In contrast to the recent studies~\cite{Trebst2022,Arun2022}, our results do not support the proposed spin-liquid states in the Heisenberg~\cite{Trebst2022}, or strongly-anisotropic ($\Delta\!=\!0.25$) nearly $XY$~\cite{Arun2022} limits. The $J_3$-width of the quantum Iz phase in the same $XY$ limit ($\Delta\!=\!0$) is also an order of magnitude narrower in our case than the one suggested in~\cite{Trebst2022}. 

While the first of the quantum phases, dZZ, missed by the previous works due to small cluster sizes or an approximate nature of their approaches~\cite{Trebst2022}, is nearly the same in the partial and full $XXZ$ models in Fig.~\ref{fig:phd}(b) and Fig.~\ref{fig:phd_full}, respectively, the Iz phase is substantially more tenuous. In fact, the initial DMRG scans have shown a direct FM-ZZ transition, with some possible narrow intermediate state. Dedicated non-scans in that region did uncover  short-range correlations in both XC and YC clusters \cite{sm}, not unlike the ones reported in Ref.~\cite{Arun2022}.  However, these spin-liquid-suspects either order on the cylinder width increase (XC), or indicate a sufficiently robust Iz order in the range of $J_3$=0.315-0.325 for $\Delta\!=\!0.25$ and $J_3$=0.34-0.36 for $\Delta\!=\!0$, see \cite{sm}.  

It is  worth noting that MAGSWT  in the $XY$ limit of the full $XXZ$ model shows a close, but  insufficient, competition of the strongly fluctuating Iz phase, rendering it absent from its version of the phase diagram in Fig.~\ref{fig:phd_full}. 

{\it Summary.}---%
In this letter, we have studied the emergence of the quantum phases that are not stable classically within a simple model of  great current interest.  We have combined  state-of-the-art  DMRG and analytical approaches to obtain  conclusive  phase diagrams of this model. It is established beyond any reasonable doubt that the  two unconventional quantum phases occupy a significant portion of this diagram, with the known phases also  extending well beyond their classical regions  and completely replacing  the less-fluctuating non-collinear phase. The results of the analytical MAGSWT approach are shown to be in a close accord with the  numerical DMRG data, providing additional insights into the energetics of the quantum stabilization of the non-classical phases and offering a systematic path for the explorations of similar models.

The proposed phase diagrams have direct relevance to a group of novel materials  and provide important guidance to the ongoing theoretical and experimental searches of the unconventional quantum states.

\begin{acknowledgments}
\emph{Acknowledgments.}---%
We are sincerely thankful to Arun Paramekanti for several important discussions and numerous exchanges. We are grateful to Ciar\'{a}n Hickey, Simon Trebst, and Yoshito Watanabe for the kind remarks and useful conversations.
The work of S.~J. and S.~R.~W. was supported by the NSF through grant DMR-2110041.
The inception of the paper, key analysis, and analytical and numerical calculations using MAGSWT by A.~L.~C. were supported by the U.S. Department of Energy, 
Office of Science, Basic Energy Sciences under Award No. DE-SC0021221.
\end{acknowledgments}

\bibliography{ref}

\end{document}


\title{Quantum Phases in the Honeycomb-Lattice $J_1$--$J_3$ Ferro-Antiferromagnetic Model: Supplemental Material}

\begin{CJK*}{UTF8}{}
\author{Shengtao Jiang (\CJKfamily{gbsn}蒋晟韬)}
\author{Steven R. White}
\author{A. L. Chernyshev}
\affiliation{Department of Physics and Astronomy, University of California, Irvine, California 92697, USA}
\date{\today}
\maketitle
\end{CJK*}

\vspace{-1.3cm}
\section{phase boundaries from the DMRG scans}
\vskip -0.3cm

\begin{figure}[h]
\centering    
\includegraphics[width=0.7\columnwidth]{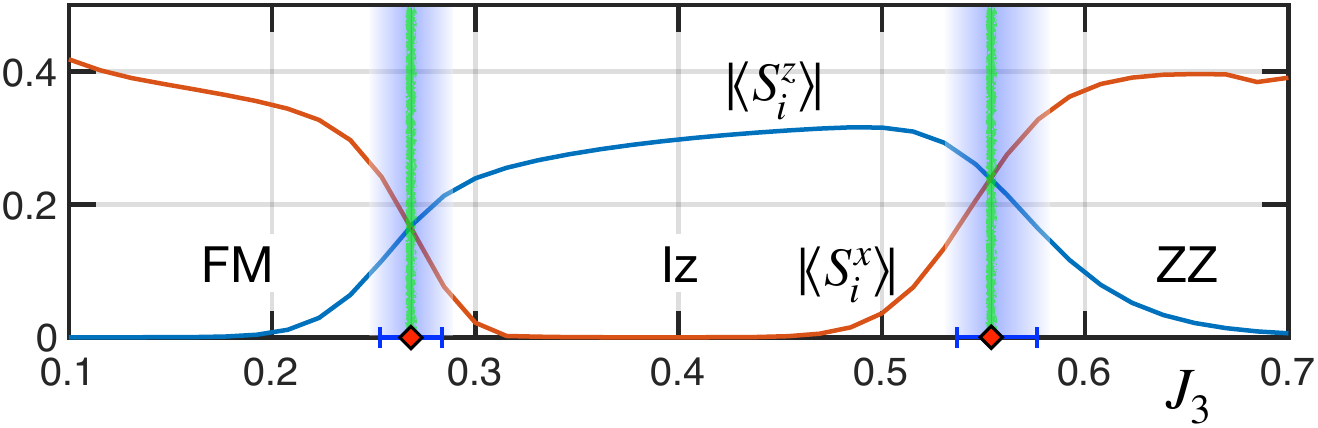}
\vskip -0.35cm
\caption{$|\langle S^z_i \rangle|$ and $|\langle S^x_i \rangle|$ along the length of a 16$\times$40 cylinder in a DMRG scan vs $J_3$ for the $\Delta\!=\!0$ limit of the $J_1^\Delta$--$J_3$ model. The crossing points are the phase boundaries and the shaded regions are the error bars.}
\vskip -0.85cm
\label{smfig:errorbar}
\end{figure}

Here we illustrate how we determine the approximate phase boundaries and corresponding error bars from the DMRG scans.  In Figure~\ref{smfig:errorbar}, we show the in-plane, $|\langle S^x_i \rangle|$, and out-of-plane, $|\langle S^z_i \rangle|$, ordered moments along the DMRG $J_3$-scan in Fig.~2(b) of the main text. Spins in the FM and ZZ phases are along the $x$ axis, while in the Iz phase they order along the $z$ axis. The transition points are chosen as the crossing points of their order parameters. Error bars are either the distance to the inflection points of the order-parameter curves or a minimum of one step of the scan (one column of the cylinder) for sharper transitions.

\section{Proximity effect in the scans and the absence of an spiral phase}
\vskip -0.5cm

\begin{figure}[h]
\centering    
\includegraphics[width=0.56\columnwidth]{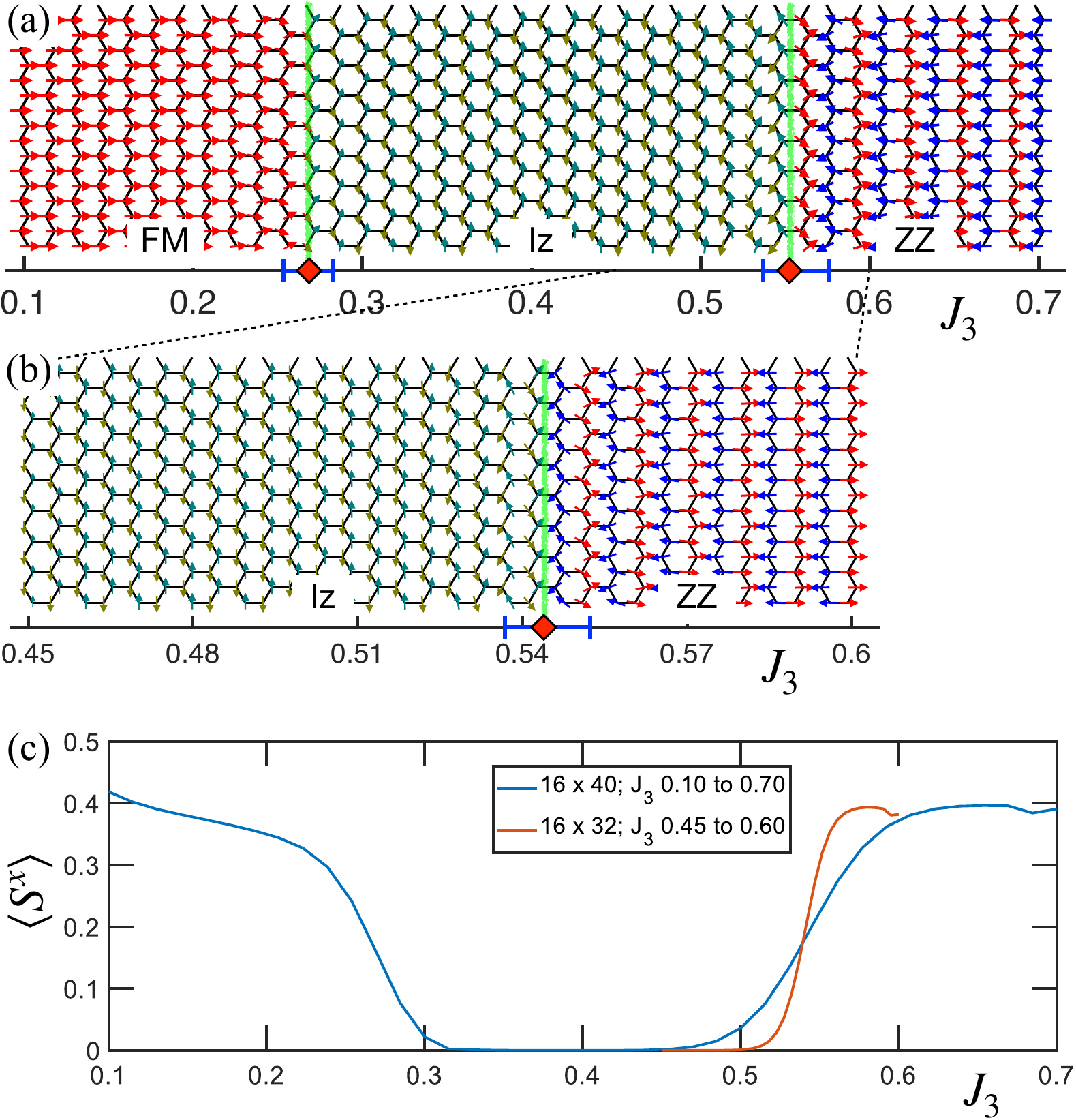}
\vskip -0.35cm
\caption{Results for the $J_1^\Delta$--$J_3$ model at $\Delta\!=\!0$. (a) The $J_3$-scan from Fig.~2(b) of the main text. (b) The ``zoom-in'' $J_3$-scan of the Iz-to-ZZ transition region in the 16$\times$32 cylinder. (c) The column-averaged $\langle S^x \rangle$ vs $J_3$ for the scans in (a) and (b).}
\label{smfig:nospiral}
\end{figure}

In some  DMRG scans, such as the one in Fig.~2(b) of the main text, reproduced in Fig.~\ref{smfig:nospiral}(a), spins at the boundary between the Iz and other phases appear to form a spiral pattern. To rule out an additional intermediate spiral phase, we perform a scan in a smaller range of the varied parameter (``zoom-in'' scan) to observe the boundary region closer. In Fig.~\ref{smfig:nospiral}(b) we focus on the transition region between the Iz phase and the ZZ phase. If the spiral phase would  exist, it would become wider in such a scan. In Fig.~\ref{smfig:nospiral}(b), the transition region has the same width (about ten columns) as in Fig.~\ref{smfig:nospiral}(a), with the transition getting sharper for the smaller gradient of $J_3$, see Fig.~\ref{smfig:nospiral}(c), strongly suggesting the absence of any intermediate phase in the thermodynamic limit. In the non-scan calculation at $J_3$=0.55 we also do not find the spiral phase. This analysis clearly shows that the spiral-like pattern in the  scans is due to a proximity effect at the phase boundary.  Similar verifications were carried out for all suspicious phases in all scans.

\section{Other DMRG scans for the partial \texorpdfstring{{\it XXZ} $J_1^\Delta$--$J_3$ model}{Lg}}
\vskip -0.3cm

\begin{figure}[h]
\centering    
\vskip -0.3cm
\includegraphics[width=0.56\columnwidth]{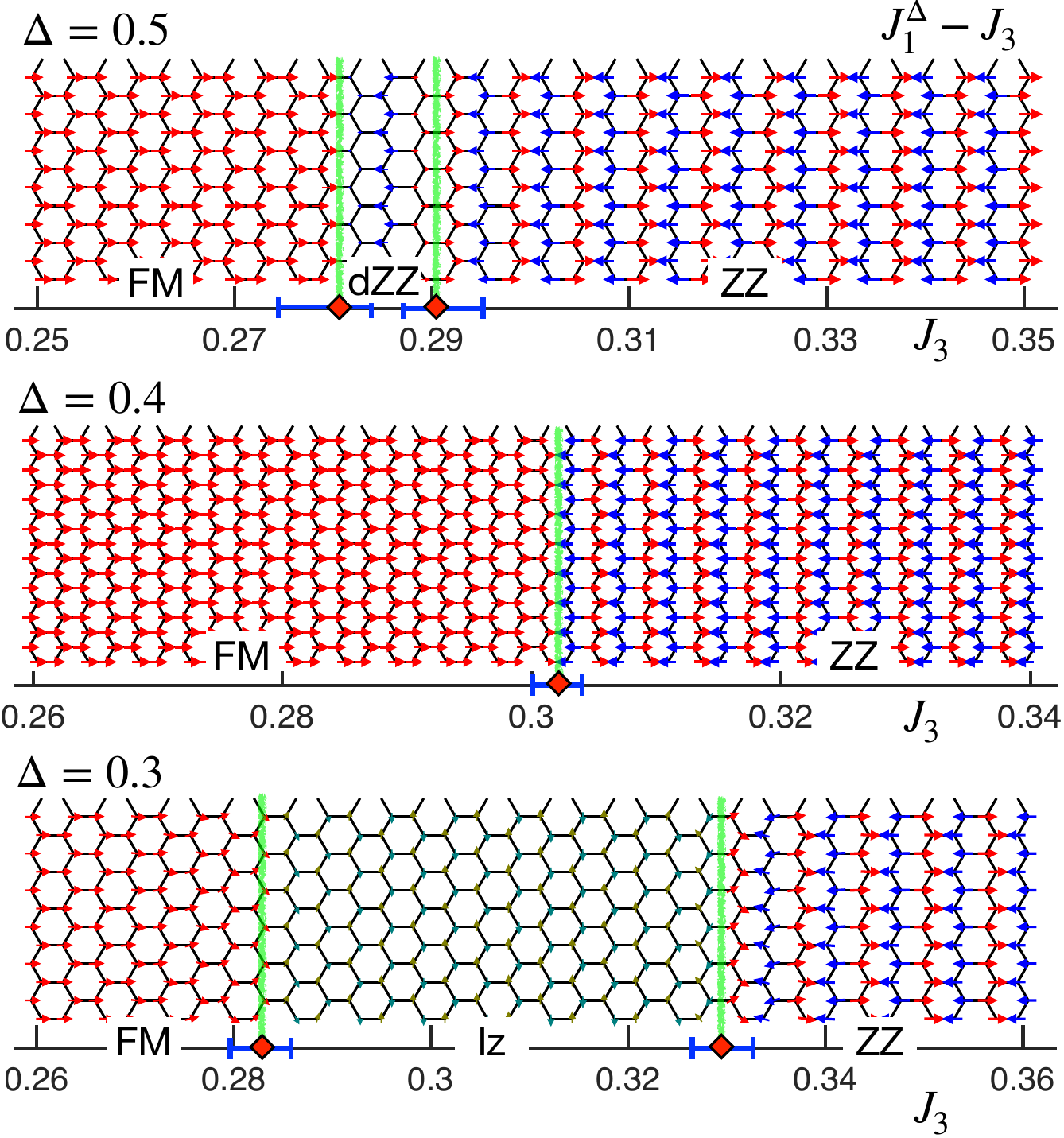}
\vskip -0.5cm
\caption{DMRG $J_3$-scans for $\Delta$=0.5, 0.4, and 0.3  in the $J_1^\Delta$--$J_3$ model. The scans for $\Delta$=0.5 and 0.3 are on the 12$\times$32 cylinders while the $\Delta$=0.4 scan is on the 16$\times$40 cylinder.}
\label{smfig:scanpxxz}
\end{figure}

In Fig.~\ref{smfig:scanpxxz}, we show additional  $J_3$-scans that are used to construct the phase diagram of the $J_1^\Delta$--$J_3$ model in Fig.~1(b) of the main text. In each scan, approximate transition boundaries with error bars are indicated.  In the $\Delta\!=\!0.5$ scan, we observe a narrow phase intervening between FM and ZZ, which is identified as the dZZ phase using non-scan calculations in the region of $J_3$ from 0.28 to 0.29 (not shown). The $\Delta\!=\!0.4$ scan in Fig.~\ref{smfig:scanpxxz} shows a direct transition from FM to ZZ.  The non-scans using smaller clusters in the vicinity of $J_3\!=\!0.3$ have initially suggested a spin-liquid (SL) state  discussed below,  which turns into ZZ order in the larger non-scan clusters. The $\Delta\!=\!0.3$ scan is similar to Fig.~2(b) of the main text with an extended region of the Iz phase intervening between FM and ZZ.

\section{DMRG Scans for the full \texorpdfstring{{\it XXZ} $J_1^\Delta$--$J_3^\Delta$ model}{Lg}}

\begin{figure}[h]
\centering    
\vskip -0.25cm
\includegraphics[width=0.7\columnwidth]{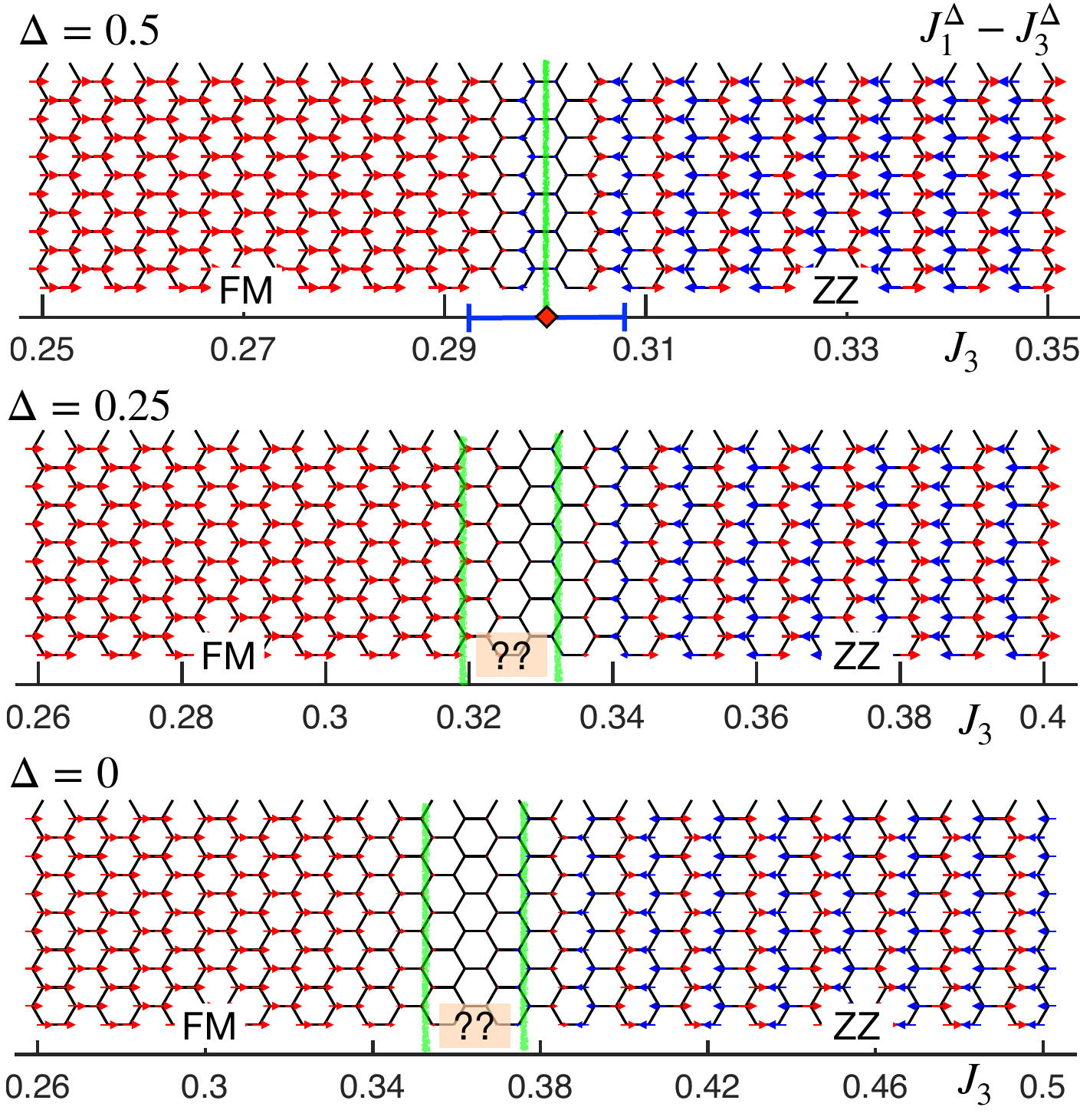}
\vskip -0.25cm
\caption{DMRG $J_3$-scans in the $J_1^\Delta$--$J_3^\Delta$ model for $\Delta$=0.5, 0.25, and 0  on the 12$\times$32 cylinders.}
\vskip -0.5cm
\label{smfig:scanfxxz}
\end{figure}

In Fig.~\ref{smfig:scanfxxz}, we show DMRG $J_3$-scans that are used to construct the phase diagram of the $J_1^\Delta$--$J_3^\Delta$ model  in Fig.~5 of the main text.  While the $\Delta\!=\!0.5$ scan looks somewhat similar to the scan for the same $\Delta$ in Fig.~\ref{smfig:scanpxxz}, it has a direct FM-ZZ transition at $J_3\!=\!0.30$, with the separate non-scan calculations showing no sign of the intermediate phase. 

In the $\Delta\!=\!0.25$ and $\Delta\!=\!0$ scans, an intermediate region is suggested with the suppressed ordered moments. As we discuss next,  initial non-scans in these regions have shown strongly anisotropic correlations, with short correlations in one direction and FM-like in the other, resembling the state that has been hypothesized as a spin liquid in Ref.~\cite{Arun2022}. Upon closer inspection and finite-size scaling, they reveal a narrow region of the Iz phase. For $\Delta\!=\!0$, $J_3\!=\!0.33$ is in the FM phase, $J_3\!=\!0.37$ is in the ZZ phase,  and $J_3\!=\!0.35$ is in the Iz phase by that analysis, confining the Iz phase between $J_3\!=\!0.34$ and 0.36.   For $\Delta\!=\!0.25$, the Iz phase is even narrower, between $J_3\!=\!0.315$ and 0.325. 

While the Iz phase in the $XY$ limit ($\Delta\!=\!0$) of the full $XXZ$ $J_1^\Delta$--$J_3^\Delta$ model  has been suggested  in Ref.~\cite{Trebst2022}, the $J_3$-width of it in our analysis is  an order of magnitude narrower than in the results of the pseudo-fermion functional renormalization group method used in Ref.~\cite{Trebst2022}.

\section{Pseudo-spin-liquid state}
\label{sec:psl}

\begin{figure}[h]
\centering    
\vskip -0.2cm
\includegraphics[width=0.7\columnwidth]{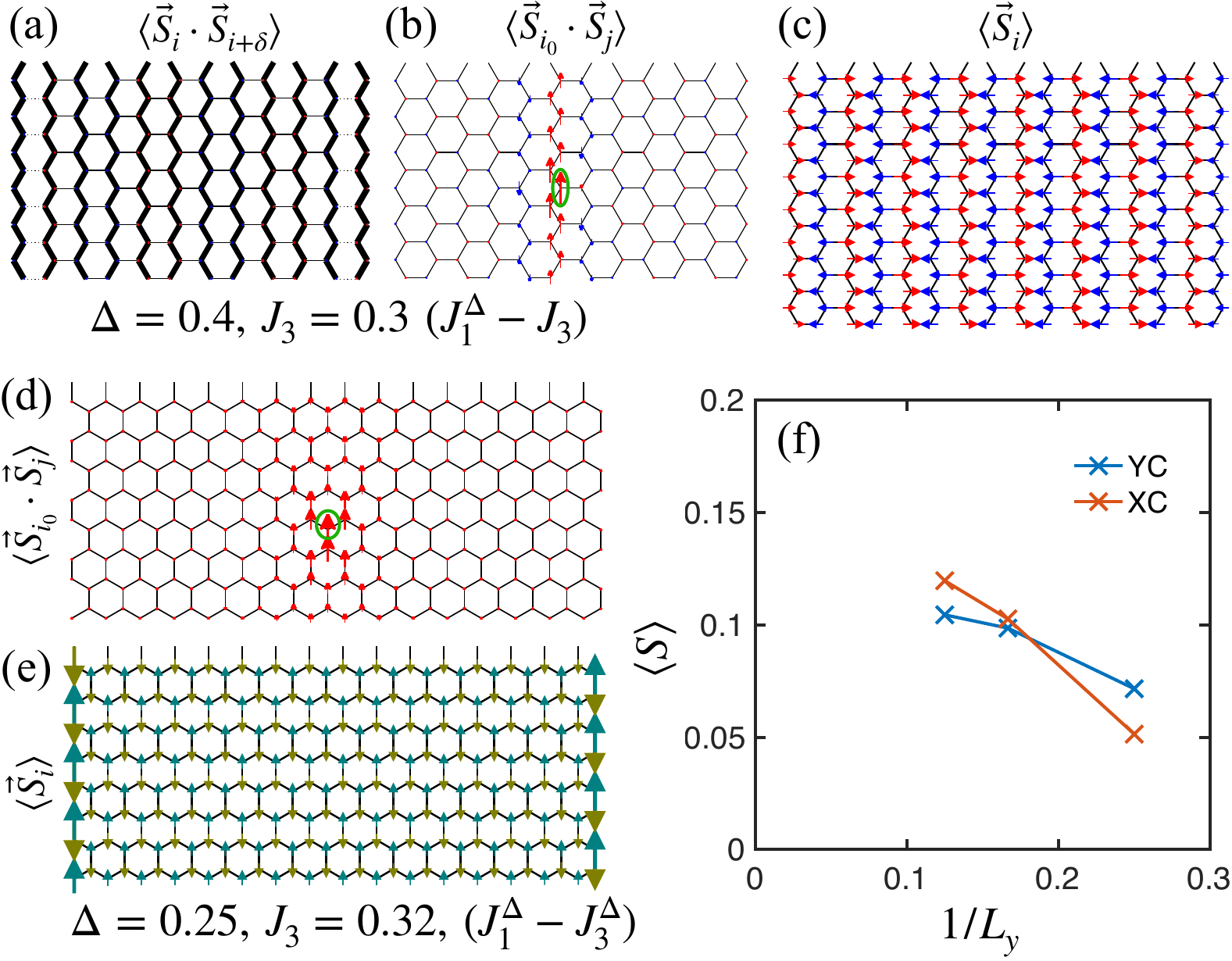}
\vskip -0.4cm
\caption{(a) and (b) 12$\times$12 XC cylinder non-scans for  $\Delta\!=\!0.4$ and $J_3\!=\!0.3$ in the $J_1^\Delta$--$J_3$ model showing: (a) nearly-zero ordered moment and nearest-neighbor $\langle \Vec{S}_i\cdot\Vec{S}_{i+\delta} \rangle$ (thickness of the bond), and (b) spin-spin correlation $\langle \Vec{S_{i_0}}\cdot\Vec{S_j} \rangle$, denoted by the length and direction of the arrow, with $i_0$ site shown by the green oval. The arrow on $i_0$ is of length 0.25. (c) Same  as  (a)  on the 16$\times$16 cylinder. (d) Same as (b) on the 8$\times$32 YC cylinder for $\Delta\!=\!0.25$ and $J_3\!=\!0.32$ in the $J_1^\Delta$--$J_3^\Delta$ model. (e) Ordered moment in (d) under the Iz pinning field of 0.5 on both edges. (f) The $1/L_y$-scaling of the Iz ordered moment in the center of the cylinder with the edge pinning fields from (e) and the XC and YC cylinders having the aspect ratio 2, which  mimics the 2D limit closely---see Ref.~\cite{FS}.}
\label{smfig:psl}
\vskip -0.4cm
\end{figure}

In some of the transition regions discussed above for both versions of the $XXZ$ $J_1$--$J_3$ model,  we have found  regimes that can be taken as  evidence for a spin-liquid state, similar to the ones reported in Ref.~\cite{Arun2022}. These include nearly zero ordered moment at  intermediate bond dimension in DMRG calculations, for which the system is expected to spontaneously break symmetry if it has an order, and the short-range spin-spin correlation in one direction, as shown in Figs.~\ref{smfig:psl}(a) and \ref{smfig:psl}(b). This anisotropy in correlations is suspicious, however, as one would expect a ``lock in'' of such 1D-like correlations into some order in a larger system. Indeed, with the increase of the system's width, one of the spin-liquid possibilities in the $J_1^\Delta$--$J_3$ model ($\Delta\!=\!0.4$), develops a ZZ order, see Fig.~\ref{smfig:psl}(c). 

Another such suspect region is in the $J_1^\Delta$--$J_3^\Delta$ model, $\Delta\!=\!0.25$, near $J_3\!=\!0.32$, similar to the one reported in Ref.~\cite{Arun2022},  but it does not follow that trend. In fact, as is shown in Fig.~\ref{smfig:psl}(d), the spin-liquid candidate looks even more realistic (less anisotropic) in the YC lattice. However, the system was tested with various boundary conditions and responded strongly to the staggered pinning field $(-1)^{i}hS^z_i$, developing a substantial Iz order, see Fig.~\ref{smfig:psl}(e), with the ordered moment nearly constant $\langle S \rangle\!\approx\!0.1$ in the bulk.  Following Ref.~\cite{FS}, we carry out an $1/L_y$-scaling of the ordered moment,  which gives a strong indication of the Iz order in the thermodynamic limit, see Fig.~\ref{smfig:psl}(f).

\section{DMRG results for the $J_1^\Delta$-$J_2^\Delta$ model}

Ref.~\cite{Trebst2022} has studied the $J_1$--$J_2$--$J_3$ $XXZ$ model, demonstrating a potentially richer structure of its phase diagram compared to the  $J_1$--$J_3$ model investigated in our work. Specifically, it was suggested that the spin-liquid phase in the isotropic Heisenberg limit is stable in a much wider region along the $J_1$--$J_2$  axis than along the $J_1$--$J_3$ axis, with a specific point $J_2\!=\!0.18$ studied in more detail. In that work, an  $XXZ$ cut of the $J_1^\Delta$--$J_2^\Delta$ model along the $\Delta$-axis for $J_2\!=\!0.18$ (and $J_3\!=\!0$) was also investigated, and a transition to an incommensurate phase from an SL phase was identified near the Heisenberg limit, at $\Delta\!=\!0.96$, with a wide range of the incommensurate phase extending down to the low values of $\Delta$. 

Here we briefly present our additional results for the $J_1^\Delta$--$J_2^\Delta$ model for this specific choice of $J_2\!=\!0.18$ and $J_3\!=\!0$, thus extending our work in a different region of the parameter space. The  summary of our results is the following. We do not find any evidence for a spin-liquid state in the Heisenberg limit of this model, and find a double-zigzag state instead. This is similar to our results for the dZZ  state in the $J_1$--$J_3$ model, found instead of the SL state suggested in Ref.~\cite{Trebst2022}, as is discussed in the main text. For the 1D phase diagram along the $\Delta$-axis for the same choice of $J_2\!=\!0.18$ and $J_3\!=\!0$, we find two transitions, one at $\Delta\!=\!0.93(2)$ and the other at $\Delta\!=\!0.86(2)$.  The lower one is a transition to a FM state, with no sign of the incommensurate phase. While the existence of a transition at $\Delta\!=\!0.93(2)$ is, ideologically, in agreement with the transition found in Ref.~\cite{Trebst2022}, in our case it is between a dZZ phase and a potentially novel triple-zigzag state that also has a significant modulation of spins, characteristic of that of the spin-density wave (SDW). We refer to it as to tZZ-SDW state.  

\begin{figure}[t]
\centering    
\includegraphics[width=0.7\columnwidth]{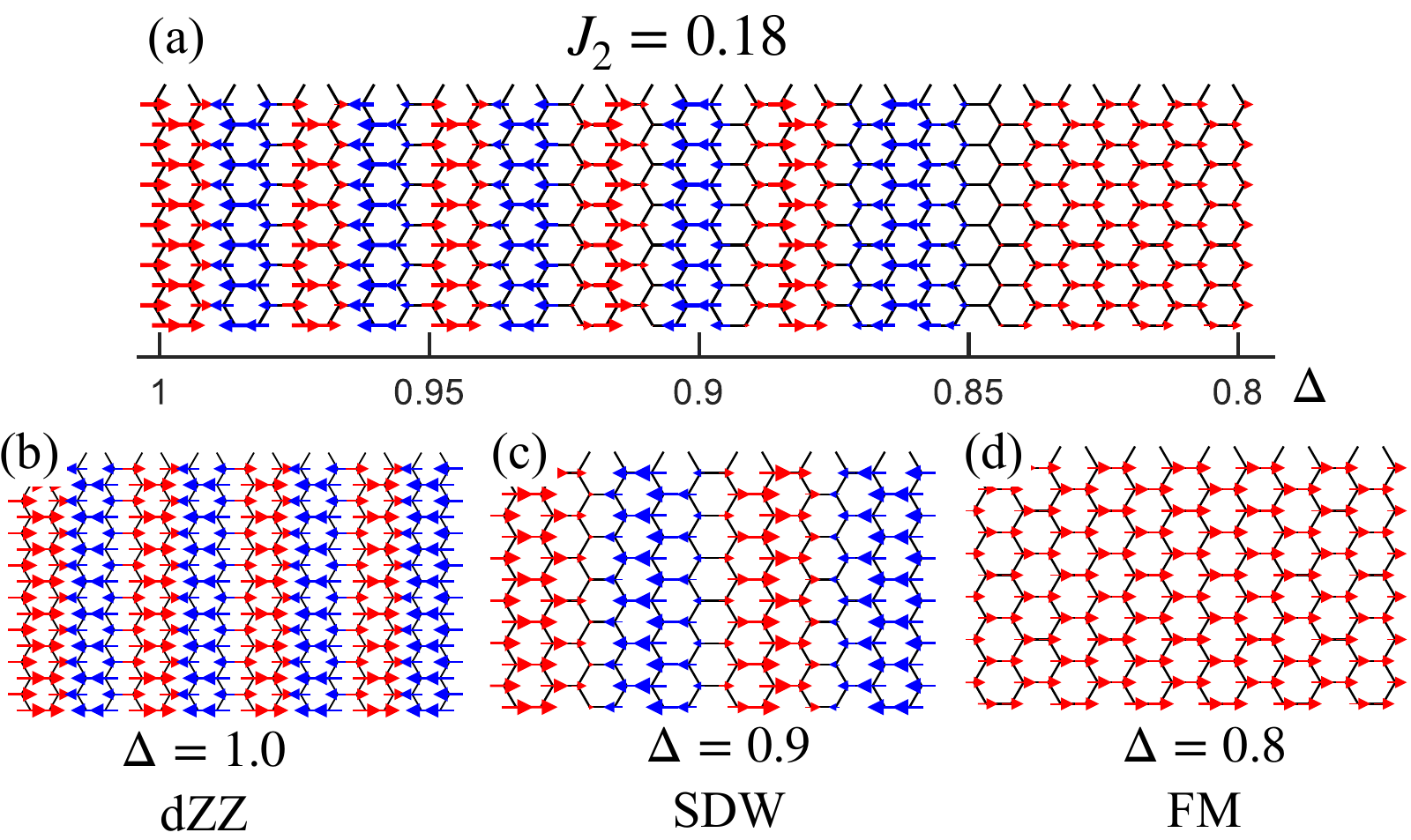}
\vskip -0.2cm
\caption{Results for the $J_1^\Delta$--$J_2^\Delta$ model for $J_2\!=\!0.18$. (a) DMRG scan on 12$\times$32 cylinder vs $\Delta$, and (b), (c), and (d) are non-scans on 16$\times$16 and  12$\times$12 cylinders for three representative values of $\Delta$.}
\label{smfig:j1j2}
\vskip -0.5cm
\end{figure}

The numerical results to substantiate these findings are presented in Fig.~\ref{smfig:j1j2}. The Fig.~\ref{smfig:j1j2}(a) part shows a scan calculation at $J_2$=0.18 vs $\Delta$ from the Heisenberg limit down to $\Delta\!=\!0.8$. The double zigzag phase at the isotropic limit ($\Delta\!=\!1.0$)  evolves into a FM state via an intermediate phase. The non-scan calculations in Fig.~\ref{smfig:j1j2}(b) and Fig.~\ref{smfig:j1j2}(d) confirm the dZZ and the FM phases at the respective ends of the scan, with both exhibiting a robust order. The non-scan for the intermediate phase at $\Delta\!=\!0.9$ in Fig.~\ref{smfig:j1j2}(c) retains the characteristics of the SDW state, as the spin's magnitude is not varied in a fashion that would be consistent with a ``simple'' triple-zigzag phase. While it is possible that the SDW variation may be an artifact of the finite cluster as the tZZ phase has a large unit cell,  the dZZ phase in the $J_1$--$J_3$ case is much more symmetric and we believe that the observed SDW variation is genuine. 

Lastly, we note that in the energy comparison for the $J_1$--$J_3$ Heisenberg case discussed in the main text and shown in Fig.~3(b), we have also investigated a stability of the triple-ZZ state. The tZZ did come very close near the FM-dZZ boundary, but did not become the ground state in that limit. In that sense, the stabilization of the tZZ phase, or a descendant of it, in a different part of the phase diagram does not come as a complete surprise.

\vspace{-0.3cm}
\section{generalized $J_1^{\Delta_1}$-$J_3^{\Delta_3}$ model for ${\rm BaCo_2(AsO_4)_2}$}

\begin{figure}[b]
\centering    
\vskip -0.5cm
\includegraphics[width=0.7\columnwidth]{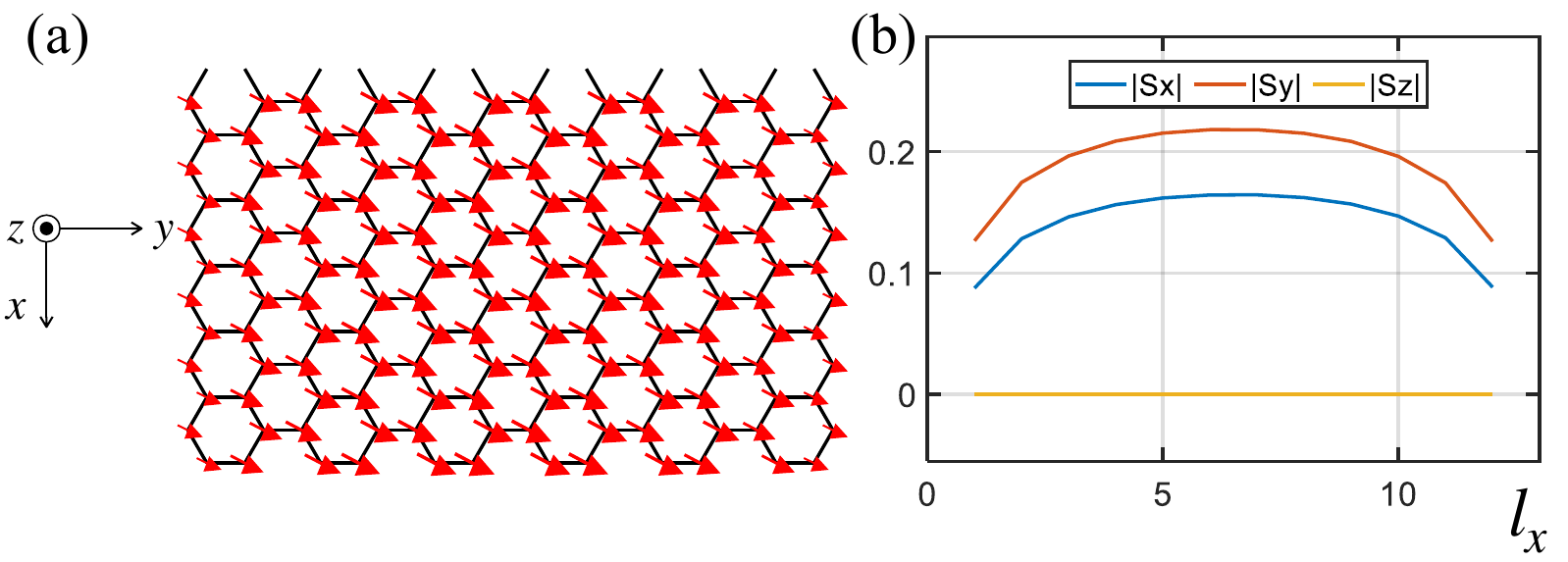}
\vskip -0.2cm
\caption{(a) Spin pattern in the 12$\times$12  DMRG cylinder and (b)  spin components in the ground state of the generalized $XXZ$ $J_1^{\Delta_1}$-$J_3^{\Delta_3}$ model model from Ref.~\cite{Broholm_2023}.}
\label{smfig:ybk}
\end{figure}

As is mentioned in the main text, extensive experimental and theoretical searches for the Kitaev magnets on the honeycomb lattice have recently expanded to the Co$^{2+}$, $S_{eff}\!=\!1/2$ materials. Among this family,  BaCo$_2$(AsO$_4$)$_2$ has received significant attention~\cite{Cava20,Regnault_18,arun-firstprinciple,pavel-dZZ,Broholm_2023}. Its minimal model description has currently coalesced to a  generalized  $XXZ$ FM-AFM $J_1^{\Delta_1}$--$J_3^{\Delta_3}$ model~\cite{arun-firstprinciple,pavel-dZZ,Broholm_2023,Arun2022} with additional Kitaev-like bond-dependent terms.

One such model parametrization was advocated in Ref.~\cite{Broholm_2023}, based on fitting experimental excitation spectrum in high fields and assuming  the spin-spiral ground state with a nearly commensurate ordering ${\bf Q}$-vector in zero field. Leaving the correctness of the latter assumption aside~\cite{Regnault_18}, the model parameters in Ref.~\cite{Broholm_2023} were constrained to match the ordering ${\bf Q}$-vector of the planar spin spiral from the {\it classical} solution of the generalized $XXZ$ $J_1^{\Delta_1}$--$J_3^{\Delta_3}$ model. 

Since we find that such a spiral state does not survive at all in the {\it quantum} $S\!=\!1/2$ version of the $XXZ$ $J_1$--$J_3$ model, as it is overtaken by the collinear phases due to quantum fluctuations, we have  checked the validity of the key assumption made in Ref.~\cite{Broholm_2023} regarding the structure of the ground state for their proposed set of parameters. The model used in Ref.~\cite{Broholm_2023} has strong  $XXZ$ anisotropies for the $J_1$ and $J_3$ terms, but of different sign,  $\Delta_1\!\approx\!0.16$ and  $\Delta_3\!=-\!0.34$, and the ratio $J_3/J_1\!\approx\!-0.33$ (see Eq.~[13] of Ref.~\cite{Broholm_2023}). The model also contains two minimal bond-dependent corrections in the $J_1$ exchange matrix.  

We have performed DMRG calculations for these parameters, including the bond-dependent terms,  on a 12$\times$12  cylinder in order to see whether the opposite sign of $\Delta_1$ and $\Delta_3$, or  the bond-dependent terms, are able to stabilize the spiral state to avoid the fate we find for it in the other models.  As is shown in Fig~\ref{smfig:ybk}, we find an FM ground state instead of the spiral state, suggesting that the model parameters for BaCo$_2$(AsO$_4$)$_2$ proposed in Ref.~\cite{Broholm_2023} are not adequate to describe its ground-state spin configuration and require a reconsideration.

\vspace{-0.3cm}
\section{Minimally augmented spin wave theory}

\vskip -0.2cm
The spin-wave approach is based on the $1/S$-expansion about a classical ground state of a spin model using bosonic representation for  spin operators \cite{hp1940}. Since the classical energy is at a minimum,  the first non-zero term of the expansion is quadratic (harmonic), yielding  the liner spin-wave theory (LSWT) Hamiltonian in a standard form
\vskip -0.15cm
\noindent
\begin{equation}
\label{eq_H_SWT}
{\cal H}=E_{cl}+\frac{1}{2}\sum_{{\bf q}} \left(\hat{\bf x}_{\bf q}^\dagger 
\hat{\bf H}_{\bf q}\hat{\bf x}_{\bf q}^{\phantom{\dagger}}
-\frac{1}{2}{\rm tr}(\hat{\bf H}_{\bf q})\right)+O(S^0),\quad\quad 
\hat{\bf H}_{\bf q}=
\left( \begin{array}{cc} 
\hat{\bf A}^{\phantom \dagger}_{\bf q} &  \hat{\bf B}^{\phantom \dagger}_{\bf q}\\[0.5ex] 
\hat{\bf B}^\dagger_{\bf q}  & \hat{\bf A}^*_{\bf -q}
\end{array}\right),
\end{equation}
\vskip -0.25cm
\noindent
where $E_{cl}$ is the classical energy, $O(S^2)$, $\hat{\bf x}^\dag_{\bf q}\!=\!\big( \hat{\bf a}^\dag_{\bf q}, 
\hat{\bf a}^{\phantom \dag}_{-{\bf q}}\big)$ is a vector of the bosonic creation and annihilation operators,  and $\hat{\bf H}_{\bf q}$ is the Hamiltonian matrix, $O(S)$, in this basis. The diagonalization of $\hat{\bf g} \hat{\bf H}_{\bf q}$, where $\hat{\bf g}$ is the diagonal para-unitary matrix, yields the LSWT magnon eigenenergies $\{\varepsilon_{1{\bf q}},  \varepsilon_{2{\bf q}}, \dots,  -\varepsilon_{1-{\bf q}},  -\varepsilon_{2-{\bf q}}, \dots\}$ \cite{Colpa} that are {\it guaranteed} to be positive definite because the expansion is around a minimum of the classical energy.

From that, the energy of the ground state to the order $O(S)$ is ${\cal E}\!=\!E_{cl}+\delta E$, where $\delta E$ is the $1/S$ quantum  correction 
\vskip -0.1cm
\noindent
\begin{equation}
\label{eq_E}
\delta E=\frac{1}{2}\sum_{\bf q} \Big(\sum_{\alpha}\varepsilon_{\alpha{\bf q}}
-{\rm tr}(\hat{\bf A}_{\bf q})\Big).
\end{equation}
\vskip -0.2cm
\noindent
When the classical state stops being a minimum as some parameter of the model is varied, the quadratic Hamiltonian in (\ref{eq_H_SWT}) ceases to be positive definite, with some of the $\varepsilon_{\alpha{\bf q}}^2$  turning negative for some momenta ${\bf q}$, and the  quantum  correction in (\ref{eq_E}) becoming ill-defined. This hinders the use of the LSWT  outside the classical region of stability of a state and limits its ability to describe the shift  of  the phase boundaries between  classical states due to quantum effects and  the appearance of the  ordered phases that are not favored classically but  stabilized in a quantum case. 

The resolution to this general conundrum that has plagued  application of the SWT to the classically unstable states was suggested in Refs.~\cite{Mila_12,Mila_13,Mila_14}.  The method consists of adding a  local field term to the Hamiltonian, $\delta{\cal H}\!=\!\mu\sum_{i} a^\dag_i a_i$ (see the main text) and referred to as the {\it minimally augmented SWT} (MAGSWT).  The minimal value of   this field is chosen from the condition that all  eigenvalues $\varepsilon_{\alpha{\bf q}}^2$  are positive definite for all the momenta ${\bf q}$.

\vspace{-0.3cm}
\subsection{LSWT for the phases of the $J_1$--$J_3$ model}

\vskip -0.2cm
The classical energies of  the collinear phases of interest per number of atomic unit cells $N_A$ are given by
\begin{align}
\label{eq_Ecl}
&E_{cl}^{\rm FM}=-3 S^2(1-J_3),\ \ E_{cl}^{\rm ZZ}=-S^2(1+3J_3),\ \ E_{cl}^{\rm Iz}=3S^2(\Delta_1-J_3\Delta_3),\ \ E_{cl}^{\rm dZZ}=-2S^2,
\end{align}
valid for any $J_3$ and $\Delta_{1(3)}$ of the model (1) of the main text, inside or outside the phase's stability region. 

Of the five phases in Fig.~1 of the main text, the magnetic unit cell in the FM and Iz phases is naturally that of the honeycomb lattice (two sites), while for the ZZ and Sp ones it can be  reduced to that by the staggered or rotated  reference frames, respectively, resulting in the $4\!\times 4$  Hamiltonian LSWT matrix $\hat{\bf H}_{\bf q}$ (\ref{eq_H_SWT}) in all four cases. For the dZZ phase, the staggered reference frame reduces the unit cell from eight to four sites   and yields the $8\!\times 8$ LSWT matrix.

The LSWT treatment of the collinear phases is rather standard and we do not elaborate on it except for a few details. 
In all two-sublattice cases, FM, ZZ, Iz, and Sp,  the LSWT matrices $\hat{\bf A}_{\bf q}$, $\hat{\bf B}_{\bf q}$ in (\ref{eq_H_SWT})  assume the same structure,  for which the  eigenvalues of the $4\!\times 4$ Hamiltonian matrix  can be found analytically. One can find additional simplifications of the eigenvalue problem for the FM and Iz phases, and in all four  cases in the  limit $\Delta_{1(3)}\!=\!0$, see also Ref.~\cite{Rastelli} for the limiting cases for the Sp phase.

In the 4-sublattice dZZ case, the eigenvalue problem for the $8\!\times 8$ matrix is not reducible to a compact analytical form. However,  analytical solutions are available for the eigenenergies at the  high-symmetry ${\bf q}\!=\!0$ and  ${\bf q}\!=\!(0,\pi/\sqrt{3})$ points in the Heisenberg limit, which are instrumental for finding the MAGSWT parameter $\mu$.

\vspace{-0.3cm}
\subsection{Finding $\mu$ in MAGSWT}

\vskip -0.2cm
In the FM, ZZ, and Iz phases, the search for the minimal value of $\mu$ for the MAGSWT  follows a similar pattern. In a simplified case, such as full $XXZ$ ($\Delta_1\!=\!\Delta_3$)  or $XY$ limits,  analytical expression for the lowest branch $\varepsilon_{1{\bf q}}^2$ simplifies sufficiently to yield the $J_3$-dependence of the offending negative minimum that needs to be lifted up by a positive shift. The required energy shift is easily related to $\mu$ with the $\Delta$-dependence of $\mu$  either absent or following trivially from the considered limiting cases. 
The resulting solutions correspond to a change  of the diagonal matrix element $A\rightarrow\bar{A}$ of the  LSWT matrix  $\hat{\bf A}_{\bf q}$, with $\bar{A}$  {\it in all three cases} given by 
\begin{align}
\label{eq_mus}
\bar{A}=A+\mu=3S\left|\bar{\gamma}_{{\bf Q}_{\rm max}}\right|, \ \  
\mbox{where} \ \ \bar{\gamma}_{\bf q}=\gamma_{\bf q}-J_3\gamma^{(3)}_{\bf q}, 
\end{align}
with the first- and third-neighbor hopping amplitudes $\gamma_\mathbf{q}=\frac{1}{3}\sum_{\alpha} e^{i\mathbf{q}{\bm \delta}_\alpha}$ and $\gamma^{(3)}_{\mathbf{q}}=\frac{1}{3}\sum_{\alpha} e^{i\mathbf{q}{\bm \delta}^{(3)}_\alpha}$,  and ${\bf Q}_{\rm max}$ defined  as
\begin{align}
\label{eq_Qm}
{\bf Q}_{\rm max}&=
\left\{\begin{array}{ll} 
(0,0), &  \ \  J_3<J_{3,c1}=0.25, \\
(Q_x,0), \ Q_x=\frac{2}{3}\cdot\arccos\left(\frac{1}{2J_3}\cdot\frac{1-3J_3}{1-2J_3}\right), & \ \  J_{3,c1}<J_3<J_{3,c2},\\
(2\pi/3,0), &  \ \ J_3>J_{3,c2}=(\sqrt{17}-1)/8\approx 0.3904,
\end{array} \right.
\end{align}
Technically, the condition for the maximum of $|\bar{\gamma}_{\bf q}|$ is related to that of the classical energy minimum in the Sp phase.

Interestingly, the resultant MAGSWT spectrum in the Iz phase and the quantum energy correction (\ref{eq_E}) that derives from it, are fully independent of the anisotropy parameters $\Delta_n$.

In the dZZ case, the search of $\mu$ has involved analysis of the spectrum obtained by a numerical diagonalization of the $\big(\hat{\bf g}\hat{\bf H}_{\bf k}\big)^2$ matrix  in the Heisenberg limit, which helped in identifying the relevant high-symmetry ${\bf q}$ points that require stabilization corrections. The diagonalization at these points can be reduced to an analytical form, which, in turn, yields the minimal value of $\mu$. In a narrow region of $0.1892\!<\!J_3\!<\!0.2030$, the two lowest unstable branches trade places and, in a row, develop negative minima at small but finite ${\bf q}$'s. For that region, we find that a straightforward linear interpolation for $\mu$ between the analytic solutions from the neighboring regions is the most effort-effective, as it stabilizes the spectrum if not with zero but with a very small gap. The resultant explicit expressions for $\mu$ are 
\begin{align}
\label{eq_mu_dZZ}
\mu&=\!
\left\{\begin{array}{ll} 
S\left(\sqrt{5\!-\!2J_3\!+\!J_3^2}\!-\!1\!-\!3J_3 \right), &   J_3\!<\!\widetilde{J}_{c1}=0.1892, \\
\mbox{interpolate}, & \widetilde{J}_{c1}\!<\!J_3\!<\!\widetilde{J}_{c2}=0.203,\\
2S\left(\sqrt{2\!-\!2J_3\!+\!J_3^2} \!-\!1\right), &  \widetilde{J}_{c2}\!<\!J_3\!<\!\widetilde{J}_{c3}=0.25,\\
2SJ_3, &  J_3\!>\!\widetilde{J}_{c3}.
\end{array} \right.
\end{align}
As in the other coplanar phases, FM and ZZ,  $\mu$ is independent of the $XXZ$ anisotropies $\Delta_n$. 

\vspace{-0.3cm}
\subsection{Energies}

\vskip -0.2cm
Following the MAGSWT strategy,  quantum corrections to the groundstate energies in all competing phases can now be calculated in a conventional $1/S$ fashion using Eq.~(\ref{eq_E}) with the expressions for the minimal chemical potential from (\ref{eq_mus}) and (\ref{eq_mu_dZZ}). Then the total energies ${\cal E}(J_3,\Delta)$ can be compared between the phases to create the phase diagram. 

\begin{figure}[b]
\vskip -0.3cm
\includegraphics[width=\linewidth]{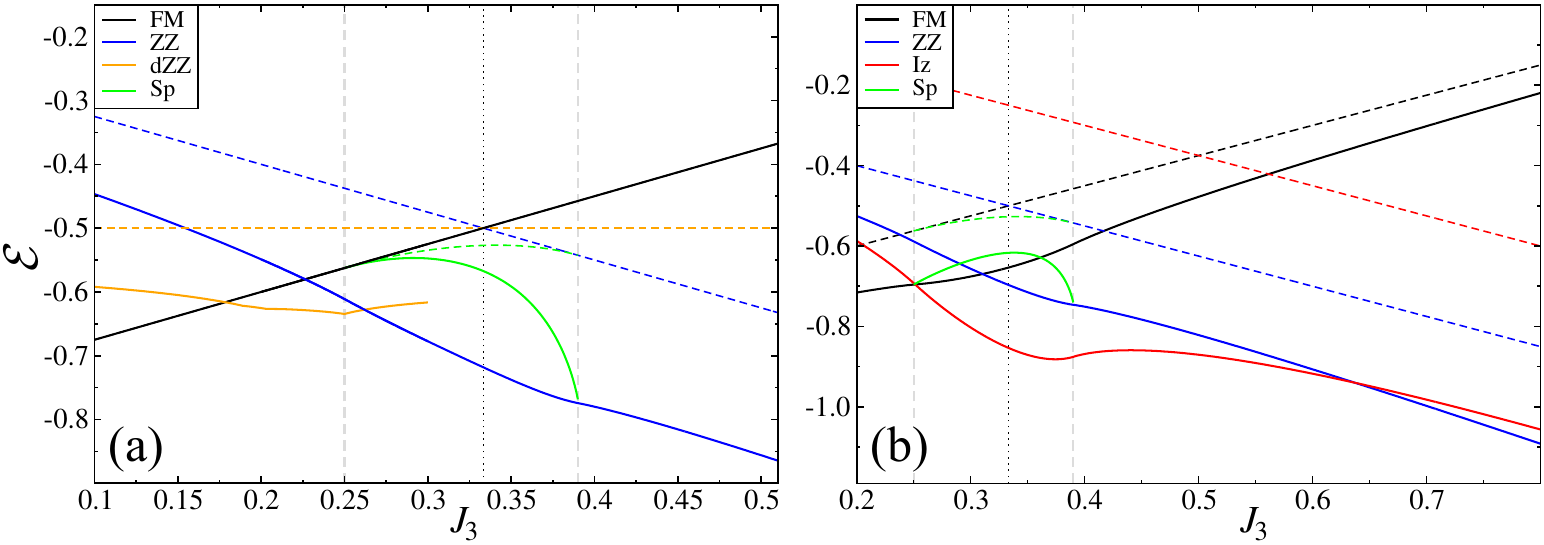}
\vskip -0.3cm
\caption{(a) The classical (dashed lines) and quantum (solid lines, from (\ref{eq_E})) energies of the FM, ZZ, Sp, and dZZ states vs $J_3$ for $\Delta_{1(3)}\!=\!1$  per atomic unit cell. The vertical dashed lines are  classical FM-Sp and Sp-ZZ  boundaries  and the dotted line is the crossing of $E_{cl}$ for the FM and ZZ states. (b) Same as (a) for the FM, ZZ, Sp, and Iz states for $\Delta_{1}\!=\!0$ and $\Delta_{3}\!=\!1$.} 
\label{Fig_Es_Heis}
\end{figure}

Figure~\ref{Fig_Es_Heis}(a) shows the $J_3$ energy-cuts  in the Heisenberg limit, $\Delta_{1(3)}\!=\!1$, of the $J_1$--$J_3$ model. The dashed lines are  classical energies  (\ref{eq_Ecl})  and solid lines are  energies with  quantum corrections (\ref{eq_E}).  The vertical dashed lines are classical FM-Sp and Sp-ZZ boundaries, $J_{3,c1}$ and $J_{3,c2}$. The dotted line is the intersection of the FM and ZZ classical energies, $J_{3}\!=\!1/3$.  The Iz phase is not competitive. The Sp phase uses standard SWT with no augmenting as it is stable through its extent. The FM is an exact eigenstate, so the  quantum corrections to it are zero. 

The first effect is the expansion of the ZZ phase (blue lines). While  the FM  is fluctuation-free, the  ZZ is not, which pushes its energy down and the crossing with the FM's energy below the $J_{3,c1}$ point where the FM  is unstable classically, superseding the non-collinear Sp phase, which is not effective in lowering its energy. However, near $J_{3,c1}$ another collinear phase, dZZ, is competitive,   making it a ground state in a finite range of $J_3$ (orange lines). 

One can note a very close agreement of the MAGSWT dZZ-ZZ transition at $J_3\!=\!0.262$ compared to the DMRG value of 0.26. On the other hand, the FM-dZZ transition is at a lower $J_3\!=\!0.1785$ than the DMRG one  at $J_3\!\approx\!0.24$. One can ascribe this difference to a larger sensitivity of the MAGSWT  phase boundaries to the higher-order corrections in this case because FM state  is non-fluctuating in the Heisenberg limit. 

For the ``partial'' $XY$ limit, with $\Delta_1\!=\!0$ and $\Delta_{3}\!=\!1$,  see Figure~\ref{Fig_Es_Heis}(b). In this case, dZZ is not competitive, but Iz  is. All phases are fluctuating in this limit, including FM. The Sp phase is not effective in benefiting from quantum fluctuations. The transition point between FM and ZZ phase is renormalized to a slightly smaller $J_3$ from its classical value. However, both are overtaken by the strongly-fluctuating Iz phase in a wide window of $J_3$. One  observation is that while the FM-Iz transition is associated with a rather steep energy crossing, the Iz-ZZ  crossing is rather shallow, suggesting stronger higher-order effects on the MAGSWT phase boundary for the latter, but not the former. This is in accord with the numerical values: $J_3\!\approx\!0.269(15)$ [DMRG] vs 0.2513 [MAGSWT] for the FM-Iz boundary and  $J_3\!\approx\!0.554(23)$ [DMRG] vs 0.637 [MAGSWT] for the Iz-ZZ boundary. Similar discrepancies for the finite $\Delta_1$ in the phase diagram in Fig.~1(b) of the main text can be attributed to the same effect.

\begin{figure}[t]
\includegraphics[width=\linewidth]{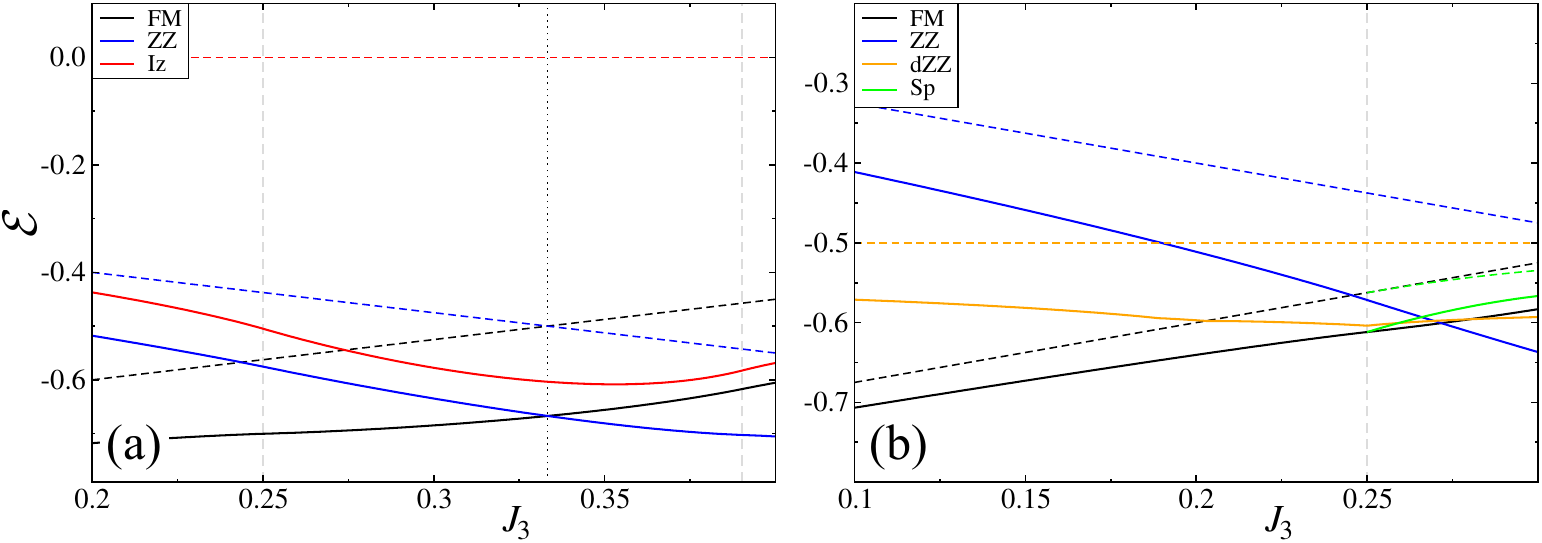}
\vskip -0.3cm
\caption{(a) Same as in Fig.~\ref{Fig_Es_Heis}(b) for the FM, ZZ, and Iz states in the ``full'' $XY$ limit, $\Delta_{1}\!=\!\Delta_{3}\!=\!0$. (b) Same as in Fig.~\ref{Fig_Es_Heis}(a) for the FM, ZZ, Sp, and dZZ states for $\Delta_{1}\!=\!0.5$ and $\Delta_{3}\!=\!1$.}
\label{Fig_Es_XYall}
\vskip -0.4cm
\end{figure}

We note that  in the ``partial'' $XY$ case in Fig.~\ref{Fig_Es_Heis}(b),  the Heisenberg $J_3$-term helps to stabilize the Iz state. The effect of the $\Delta_3$ anisotropy  is tested by the ``full'' $XY$ limit of the model, in which the  benefit of the out-of-plane spin-coupling is absent. The $J_3$-cut in this limit is shown in Figure~\ref{Fig_Es_XYall}(a). The Iz phase can be seen as remarkably effective at lowering its energy, with the quantum fluctuation part being about four times of that for the FM and ZZ states. However, while being closely competitive, the Iz phase is not stable in the full $XY$ limit according to MAGSWT.  This result is, superficially,  in a disagreement with the  DMRG, which does show a narrow strip of the Iz phase in Fig.~5 of the main text. Nevertheless, with the energy curves in Fig.~\ref{Fig_Es_XYall}(a) and Fig.~\ref{Fig_Es_Heis}(b) in mind, it is clear that the MAGSWT misses Iz phase in the full  $XY$ limit only  slightly.  

An additional $J_3$-cut for $\Delta_1\!=\!0.5$ and Heisenberg $J_3$ is shown in Fig.~\ref{Fig_Es_XYall}(b). Here, the competing phases are the same as in Fig.~\ref{Fig_Es_Heis}(a), with the dZZ phase coming extremely close, but not able to stabilize, yielding a direct FM-ZZ transition for this value of $\Delta_{1}$. This is in a close agreement with DMRG, which shows a narrow dZZ slice for $J_3$ between 0.280(4) and 0.290(6)  at this $\Delta_{1}$, with the FM-ZZ transition being direct for the next cut at $\Delta_1\!=\!0.4$, see Fig.~1(b) of the main text. Given the  energy differences in Fig.~\ref{Fig_Es_XYall}(b), the agreement is indeed very close. 

Such additional insights into the energetics of the competing phases are instrumental for the understanding of their competition. They also underscore the undeniable  success of the MAGSWT  in describing classically unstable states. 

\bibliography{ref}